\documentclass[11pt,a4paper]{article}	
\usepackage{fullpage}
\usepackage{amsmath}
\usepackage{amsbsy}
\usepackage{amsfonts}
\usepackage{graphicx}
\usepackage{dsfont}
\newcommand{\be}{\begin{equation}}
\newcommand{\ee}{\end{equation}}
\newcommand{\hmin}{H_\mathrm{min}}
\newcommand{\tr}{\mathrm{tr}}
\newcommand{\id}{\mathbb{I}}
\newcommand{\x}{x^*}
\newcommand{\y}{y^*}

\usepackage{footnote} 
\usepackage{epstopdf}

\begin{document}

\title{Using complete measurement statistics for \\ optimal device-independent randomness evaluation}

\author{O. Nieto-Silleras, S. Pironio, J. Silman\\
\small Laboratoire d'Information Quantique, Universit\'{e} Libre de Bruxelles (ULB), Bruxelles, Belgium}

\maketitle
\begin{abstract}
The majority of recent works investigating the link between non-locality and randomness, e.g. in the context of device-independent cryptography, do so with respect to some specific Bell inequality, usually the CHSH inequality. However, the joint probabilities characterizing the measurement outcomes of a Bell test are richer than just the degree of violation of a single Bell inequality. In this work we show how to take this extra information into account in a systematic manner in order to optimally evaluate the randomness that can be certified from non-local correlations. We further show that taking into account the complete set of outcome probabilities is equivalent to optimizing over all possible Bell inequalities, thereby allowing us to determine the optimal Bell inequality for certifying the maximal amount of randomness from a given set of non-local correlations.
\end{abstract}

\section{Introduction}
In the context of any non-signaling theory, and in particular in the context of quantum theory, outcomes of measurements on separate systems leading to a Bell violation cannot be completely pre-determined, i.e. the violation of a Bell inequality guarantees the presence of genuine randomness. This link between non-locality \cite{rmp} and randomness is interesting on the  fundamental level \cite{Valentini,Acin12}, but is also the main ingredient behind device-independent randomness generation (DIRG) \cite{Colbeck07,Pironio10A,Pironio13A,Fehr13,Vazirani12A}, randomness amplification \cite{Colbeck12,Gallego12}, and device-independent quantum key distribution (DIQKD) \cite{Barrett05,Mayers04,Acin07,Masanes11,Renner,Pironio13B,Vazirani12B}.

At the basis of such developments lies a quantitative relation between the amount of randomness that is necessarily produced in a Bell experiment and the degree of violation of a certain Bell inequality, such as the CHSH inequality \cite{chsh,Pironio10A}, the chained inequality \cite{chained,Barrett05,BHP,Colbeck12}, or a Mermin-type inequality \cite{mermin,Colbeck07,Gallego12}. However, the set of data obtained in a Bell experiment is much richer than just the value of the  violation of some Bell inequality. For example, in a CHSH experiment there are eight independent probabilities that determine the single number corresponding to the amount of CHSH violation. Moreover, in \cite{Acin12} it was shown that there exist two-input two-output Bell inequalities that can allow for the certification of more randomness than the CHSH inequality. Similar examples have been provided in \cite{marcin}. Such results imply that taking into account extra data beyond the value of a single Bell violation can be useful, but leave open the questions of just how useful and how to do so in a systematic manner. 

These questions are especially relevant now that the detection loophole has been closed (albeit re-opening the locality loophole) with entangled photons \cite{Giustina12,Christensen13}, opening the door for high rate DIRG. Nevertheless, there is still work to be done on the theoretical level before we can realize this goal efficiently. In particular, low detection efficiencies ($\sim 0.75$) necessitate using states of  low entanglement (for efficiencies below $\simeq 0.82$ the CHSH inequality cannot be violated using maximally entangled two-qubit states \cite{Eberhard93}), for which  the CHSH inequality is not optimal with respect to randomness certification \cite{Acin12}.

In this work we show how to evaluate the randomness produced in a Bell test, or, more specifically, how to obtain the device-independent guessing probability (DIGP) by systematically taking into account the complete non-local behavior, rather than just the violation of some pre-specified Bell inequality. We also show that for any set of non-local correlations, there exists a Bell inequality that is optimal for certifying the maximal amount of randomness given these correlations. Regarding this, we note that while the protocols in \cite{Pironio10A,Pironio13A,Fehr13,Masanes11,Renner,Pironio13B} are general in the sense that they are not formulated with respect to some specific Bell inequality, they do not tell us the optimal Bell inequality to use given the measurement data. We then show how the optimal value of the DIGP and the associated optimal Bell inequality can be computed using the semidefinite programming (SDP) hierarchy introduced in \cite{Navascues07}. 
Finally, we study three numerical examples illustrating the advantage in taking into account the complete non-local behavior, as opposed to taking into account only the violation of a specific Bell inequality.

\section{Background: the device-independent guessing probability}
We consider the following setting. Alice has access to a pair of quantum devices, or boxes, $\mathcal{A}$ and $\mathcal{B}$, which she can prevent from communicating at will, and whose internal state may be correlated with a system in the possession of an adversary Eve (or equivalently to the environment). The joint state of the boxes and Eve's system is described by a quantum state $\rho_{\mathcal{ABE}}\in \mathcal{H_A}\otimes \mathcal{H_B}\otimes \mathcal{H_E}$. Alice introduces inputs $x$ and $y$, each chosen at random from the finite set $\{1,\,\ldots,\,n\}$ into boxes $\mathcal{A}$ and $\mathcal{B}$ and obtains outputs $a$ and $b$, respectively, each taking one of the values $\{1,\,\ldots,\,d\}$. This process is described by a pair of POVMs with elements $\{ M_{a \mid x}\}$ and $\{ M_{b \mid y} \}$, each acting on $\mathcal{H_A}$ and $\mathcal{H_B}$, respectively. The joint probability that the outputs $a$ and $b$ are obtained given the inputs $x$ and $y$ is $p_{\mathcal{AB}}(ab|xy)=\tr\left(\rho_{\mathcal{AB}} M_{a|x}\otimes M_{b|y}\right)$, where $\rho_\mathcal{AB}=\tr_\mathcal{E}\left(\rho_\mathcal{ABE}\right)$. There are a total of $d^2n^2$ such joint probabilities, which we view as the components of a vector $\mathbf{p}=\{p_{\mathcal{AB}}(ab|xy)\}\in\mathbb{R}^{d^2n^2}$. We refer to this vector as the (non-local) \emph{behavior} characterizing Alice's devices.

We refer to a specific state $\rho_{\mathcal{ABE}}$ and sets of measurement operators $\{M_{a \mid x}\}$ and $\{M_{b\mid y} \}$, yielding the behavior $\mathbf{p}$, as a quantum realization $Q$ of $\mathbf{p}$. 
We denote by $\mathcal{Q}$ the convex set of all behaviors $\mathbf{p}\in\mathbb{R}^{d^2n^2}$ that admit a valid quantum realization $Q$. In the following, it will be useful to consider measurements on unnormalized quantum states $\tilde \rho_{\mathcal{AB}}$ (i.e. $\text{tr}\left(\tilde \rho_{\mathcal{AB}}\right)\geq 0$). We denote the corresponding behaviors by $\mathbf{\tilde p}$ and define their norm as $\text{tr}\left(\mathbf{\tilde p}\right)=\text{tr}\left(\tilde \rho_{\mathcal{AB}}\right)$. We denote by $\mathcal{\tilde Q}$ the corresponding set of unnormalized quantum behaviors, which is a convex cone.

In general, different quantum realizations $Q$ are possible for a given behavior $\mathbf{p}$. Our aim is to quantify the randomness generated by the boxes from $\mathbf{p}$ alone, independently of the possible underlying quantum realizations $Q$ compatible with $\mathbf{p}$. To simplify the notation, we describe in the following how to quantify the \emph{local} randomness associated with box $\mathcal{A}$'s output $a$ when a certain input $x=\x$ is used.   The \emph{global} randomness associated  with both boxes' outputs $a$ and $b$ for a given pair of inputs $x=\x$ and $y=\y$ can be treated analogously.

To begin, let us fix a specific quantum realization $Q$ compatible with $\mathbf{p}$. This quantum realization defines an initial state $\rho_{\mathcal{ABE}}$ and  sets of projectors $\{ M_{a\mid x} \}$ and $\{ M_{b\mid y} \}$ \footnotemark[1] \footnotetext[1]{We can always restrict to projectors by increasing the dimension of the Hilbert space. No loss of generality will be incurred by this, since we will be working in device-independent settings.}.  After Alice's measurement the correlations between her classical output $a$ and the quantum information held by Eve are described by the classical-quantum state $\sum_a p_\mathcal{A}(a|\x)|a\rangle\langle a|\otimes \rho_{\mathcal{E}}^{a\x}$, where $\rho_{\mathcal{E}}^{a\x}$ is the reduced state of Eve given that Alice performed measurement $\x$ and obtained outcome $a$.
The randomness of box $\mathcal{A}$'s output given this side information can be quantified by the guessing probability \cite{Konig09,Acin12}:  the average probability that Eve  correctly guesses box $\mathcal{A}$'s output using an optimal strategy. Such an optimal strategy is described by a $d$-element POVM $\{M_{a|z}\}$ that Eve performs on her system; if she obtains the output $a$, which happens with probability $p_\mathcal{E}(a|z,\,a',\,\x,\,Q)=\tr\bigl(\rho_\mathcal{E}^{a'\x}M_{a \mid z}\bigr)$ when her system is in the reduced state $\rho_{\mathcal{E}}^{a'\x}$, she guesses that box $\mathcal{A}$'s output was $a$. Optimizing over all possible measurements, her average probability of guessing correctly is thus given by
\be \label{guessprobq}
G(A|E,\,\x,\,Q)=\max_{\{M_{a|z}\}} \sum_{a} p_\mathcal{A}(a|\x,\,Q)p_\mathcal{E}(a|z,\,a,\,\x,\,Q)\,.
\ee
The above expression defines the guessing probability, which is related to the quantum min-entropy $\hmin(A|E,\,\x,\,Q)$ through $G(A|E,\,\x,\,Q)=2^{-\hmin(A|E,\,\x,\,Q)}$ \cite{Konig09} \footnote{The guessing probability or equivalently the min-entropy is an operational measure of randomness: if $\rho_{\mathcal{KE}}=\sum_{k=1}^{d} p(k)|k\rangle\langle k|\otimes \rho_{\mathcal{E}}^{k}$ is a cq-state with guessing probability $G(K|E)\leq 2^{-t}$, then a randomness extractor can be used to map $k\in\{1,\ldots,d\}$ to a $t$-bit string $k'\in\{1,\ldots,2^t\}$ that is close to being uniformly random and uncorrelated to the adversary, that is $\rho_{\mathcal{K}'\mathcal{E}}$ is close in trace-distance to the state $\sum_{k'=1}^{2^{t}} 2^{-t}|k'\rangle\langle k'|\otimes \sigma_{\mathcal{E}}$ \cite{Konig09}.}. Note that in the above definition we made the dependence on $Q$ explicit to stress that we are considering a given quantum realization $Q$. Since our aim is to obtain a bound on the randomness of the outputs that depends only on $\mathbf{p}$, but not on a specific quantum realization $Q$ of $\mathbf{p}$, we must further maximize $G(A|E,\,\x,\,Q)$ over all $Q$ compatible with $\mathbf{p}$:
\be \label{guessprob}
G(A|E,\,\x)
=\max_{Q,\, \{M_{a|z}\}} \sum_{a} p_\mathcal{A}(a|\x,\,Q)p_\mathcal{E}(a|z,\,a,\,\x,\,Q)\,.
\ee
This defines the DIGP, the quantity which interests us in this work.

\section{The device-independent guessing probability as a conic linear program}
We have expressed the guessing probability as an average over Eve's probabilities conditioned on box $\mathcal{A}$'s outputs, but we can also express it, using Bayes' rule, as an average over Alice's probabilities conditioned on Eve's outcomes:
\be \label{guessprob2}
G(A|E,\,\x)=\max_{Q,\,\{M_{a|z}\}} \sum_a p_\mathcal{E}(a|z,\,Q)p_\mathcal{A}(a|\x,\,a,\,z,\,Q)\,.
\ee
Here $p_\mathcal{E}(a|z,\,Q)$ is the probability that Eve obtains the outcome $a$ and $p_\mathcal{A}(a'|\x,\,a,\,z,\,Q)$ is the probability that box $\mathcal{A}$ outputs $a'$ conditioned on that event. More generally, conditioning on Eve's outcomes defines a family of behaviors $\mathbf{p}^{azQ}$ for boxes $\mathcal{A}$ and $\mathcal{B}$, or more conveniently of unnormalized behaviors $\mathbf{\tilde p}^{azQ}=p_\mathcal{E}(a|z,\,Q)\mathbf{p}^{azQ}\in \mathcal{\tilde Q}$. Note that averaging over these behaviors yields back the given behavior characterizing the boxes: $\sum_a \mathbf{\tilde p}^{azQ}=\mathbf{p}$.  Every choice of $Q$ and $\{M_{a|z}\}$ defines a family of quantum behaviors satisfying this property. Conversely, it is not difficult to see that any set of behaviors $\mathbf{\tilde p}^{a} \in \mathcal{\tilde Q}$ satisfying $\sum_a \mathbf{\tilde p}^{a}=\mathbf{p}$ can be interpreted as describing the  conditional joint output probabilities of boxes $\mathcal{A}$ and $\mathcal{B}$  for some quantum realization $Q$ and POVM $\{M_{a\mid z} \}$ performed by Eve. In terms of the unnormalized behaviors, we can write Eq. (\ref{guessprob2}) as $G(A|E,\,\x)
=\max_{Q,\,\{M_{a|z}\}} \sum_{a} \tilde p_\mathcal{A}(a|\x,\,a,\,z,\,Q)$ and thus the DIGP associated with $\mathbf{p}$ is the solution to the following optimization problem 
\begin{eqnarray}
G(A|E,\,\x)  = & \displaystyle\max_{\{\mathbf{\tilde p}^{a}\}}&\sum_a \tilde p^a(a|\x)\label{optim}\\
& \mathrm{s.t.}& \sum_{a} \mathbf{\tilde p}^{a}=\mathbf{p}\,,\; \; \mathbf{\tilde p}^{a}\in \mathcal{\tilde Q}\,,\; \; a=1,\ldots,\,d\,,\nonumber\end{eqnarray}
where the $\mathbf{\tilde p}^a$s are the optimization variables. This is a typical instance of a conic linear program \cite{Boyd04}, i.e. the optimization of a linear objective function ($\sum_a \tilde p^a(a|\x)$) subject to linear constraints $(\sum_{a} \mathbf{\tilde p}^{a}=\mathbf{p})$ and to the constraint that the optimization variables belong to a convex cone (the constraints $\mathbf{\tilde p}^{a}\in \mathcal{\tilde Q}$, since $\mathcal{\tilde Q}$ is a closed convex cone).

The program Eq. (\ref{optim}) has a simple physical interpretation. Any feasible point corresponds to a possible quantum decomposition $\mathbf{p}=\sum_a \mathbf{\tilde p}^a$ of the behavior $\mathbf{p}$. From the point of view of an adversary, such a decomposition can be understood as a strategy where with probability $\text{tr}\left(\mathbf{\tilde p}^a\right)$ the adversary guesses that box $\mathcal{A}$'s output was $a$ and prepares the quantum behavior $\mathbf{p}^a=\mathbf{\tilde p}^a/\text{tr}\left(\mathbf{\tilde p}^a\right)$. The probability of correctly guessing box $\mathcal{A}$'s output in this strategy is $\sum_a \tilde p^a(a|\x)$. The program Eq. (\ref{optim}) simply searches for the optimal quantum strategy that maximizes this expression. 

\section{Dual formulation and optimal Bell expressions}

Every conic linear program admits a dual formulation (see, e.g. \cite{Boyd04}), which in the case of Eq. (\ref{optim}) is readily seen to be	
\begin{eqnarray} \label{optimdual}
D(A|E,\x)  = & \displaystyle\min_{\mathbf{f}}&\textbf{f}\cdot\textbf{p} \\
& \mathrm{s.t.} &    p'(a|\x)\leq_\mathcal{Q} \textbf{f}\cdot\textbf{p}'\,, \quad a=1,\ldots,d\,.\nonumber\end{eqnarray}
In the above problem the optimization variable is the vector $\textbf{f}\in \mathbb{R}^{d^2n^2}$. It can be interpreted as defining a Bell expression whose expectation value is $\textbf{f}\cdot\textbf{p}=\sum_{a,\,b,\,x,\,y}f_{abxy}p_{\mathcal{AB}}(ab|xy)$. That is, it defines a linear form in the behavior $\mathbf{p}$. 
The constraint $p'(a|\x)\leq_\mathcal{Q} \textbf{f}\cdot\textbf{p}'$ means that $p'(a|\x)\leq \textbf{f}\cdot\textbf{p}'$ should hold for all $\mathbf{p}'\in \mathcal{Q}$. Whenever $\mathbf{f}$ satisfies this constraint, the expectation value $\textbf{f}\cdot\textbf{p}$ provides an upper-bound on the guessing probability since
\begin{eqnarray}\label{bound}
G(A|E,\,\x)&=&\max_{\{\mathbf{\tilde p}^{a}\}} \sum_a \tilde p^{a}(a|\x)\nonumber\\
&\leq & \max_{\{\mathbf{\tilde p}^{a}\}} \sum_a \mathbf{f}\cdot{\mathbf{\tilde p}}^{a}\\
&=&\max_{\{\mathbf{\tilde p}^{a}\}} \mathbf{f}\cdot \Bigl(\sum_a \mathbf{\tilde p}^{a}\Bigr)=\mathbf{f}\cdot \mathbf{p}\nonumber\,.
\end{eqnarray}
In particular, given a fixed Bell expression, such as the CHSH expression $\mathbf{c}$, one can determine coefficients $\alpha$ and $\beta$ (effectively defining a new linear form $\mathbf{f}=\alpha\, \mathbf{c}+\beta$) such that $p(a|\x)\leq_\mathcal{Q} \alpha\,\textbf{c}\cdot\textbf{p}+\beta$ and thus $G(A|E,\x)\leq \alpha\,\mathbf{c}\cdot\mathbf{p}+\beta$.
Such bounds on the DIGP are the ones that are used in most works related to DIRG or DIQKD, see e.g. \cite{Pironio10A,Pironio13A,Fehr13,Vazirani12A,Silman13} and \cite{Masanes11,Renner,Pironio13B,Vazirani12B}, respectively. 
The program Eq. (\ref{optimdual}) goes further since it does not assume a fixed Bell expression, but determines the linear form that yields the lowest upper-bound $D(A|E,\,\x)$ on the DIGP for a given behavior $\mathbf{p}$. 

The fact that the dual optimal solution $D(A|E,\,\x)\geq G(A|E,\,\x)$ yields an upper bound on the primal optimal solution is a general result that holds between any primal and dual conic linear program pairs. Provided that one of the two programs admits a strictly feasible solution, it further holds that there is no gap between the primal and dual optimal solutions, i.e. $G(A|E,\,\x)=D(A|E,\,\x)$. This is the case here since the form $\mathbf{f}$, defined by $f_{abxy}=1$ for all $a,\,b,\,x$, and $y$, satisfies $\mathbf{f}\cdot \mathbf{p}=n^2$, and consequently $p(a|\x)<_\mathcal{Q}\mathbf{f}\cdot \mathbf{p}$, and so represents a strictly feasible point of the dual problem.

The programs Eqs. (\ref{optim}) and (\ref{optimdual}) are equivalent but have different interpretations. As we have explained above, the feasible points of the primal program correspond to explicit strategies for the adversary. Any such strategy yields a lower-bound on the DIGP. The primal program Eq. (\ref{optim}) searches for the optimal strategy that maximizes the guessing probability. On the other hand,  any feasible point of the dual program corresponds to a Bell expression, which certifies that a certain amount of randomness is present in the given behavior $\mathbf{p}$, and yields an upper-bound on the DIGP.  The dual program Eq. (\ref{optimdual}) searches for the Bell expression which certifies the maximal amount of randomness. The duality theorem of conic linear programming tells us that the optimal solutions of both programs are identical, and thus that for every behavior $\mathbf{p}$ there exists a Bell expression, which certifies the full amount of randomness present in the correlations.

\section{Semidefinite programming relaxations}
The above conic linear programming formulations of the DIGP are in general difficult to implement exactly. However, they can be relaxed using the SDP method introduced in \cite{Navascues07,Navascues08}. This method introduces a hierarchy of convex sets $\mathcal{\tilde Q}_1\supseteq \mathcal{\tilde Q}_2 \supseteq \ldots\supseteq \mathcal{\tilde Q}$, which approximate the quantum set $\mathcal{\tilde Q}$ from the outside \footnotemark[2] \footnotetext[2]{The hierarchy as presented in \cite{Navascues07,Navascues08} applies to normalized behaviors $\mathbf{p}\in\mathcal{Q}$, but it can be can be trivially adapted to the unnormalized behaviors $\mathbf{\tilde p}\in\mathcal{\tilde Q}$ by removing the normalization constraint, e.g. $\Gamma_{11}=1$ in the notation of \cite{Navascues07,Navascues08}.}. The hierarchy of programs 
\begin{eqnarray}\label{sdp}
G_k(A|E,\,\x)  = & \displaystyle\max_{\{\mathbf{\tilde p}^{a}\}}& \sum_a \tilde p^a(a|\x)\\
& \mathrm{s.t.} &  \sum_{a} \mathbf{\tilde p}^{a}=\mathbf{p}\,,\; \;
 \mathbf{\tilde p}^{a}\in \mathcal{\tilde Q}_k\,,\; \; a=1,\ldots,d\nonumber\end{eqnarray}
therefore provides a sequence of relaxations to Eq. (\ref{optim}), which yields 
upper-bounds $G_1(A|E,\,\x)\geq G_2(A|E,\,\x)\geq \ldots \geq G(A|E,\,\x)$ on the DIGP. In this approach 
a behavior $\mathbf{\tilde p}$ belongs to $\mathcal{\tilde Q}_k$ if and only if there exists a positive semidefinite matrix $\Gamma_k\succeq 0$ satisfying a series of linear constraints of the form $\text{tr}\left(G\, \Gamma_k\right)=\mathbf{h}\cdot\mathbf{\tilde p}$ (see \cite{Navascues08,Pironio10B} for details). Since the objective function and the first set of constraints in Eq. (\ref{sdp}) are also linear, the problems Eq. (\ref{sdp}) can be cast as SDP problems for which efficient algorithms are available.

This SDP hierarchy can also be understood from the perspective of the dual problem Eq. (\ref{optimdual}). To see this, we note that the constraint $p'(a|\x)\leq_\mathcal{Q} \textbf{f}\cdot\textbf{p}'$ in Eq. (\ref{optimdual}) is equivalent to $\langle \psi|\mathcal{F}_a|\psi\rangle\geq 0$ for all possible quantum states $|\psi\rangle$ and all possible $\mathcal{F}_a$ of the form $\mathcal{F}_a=\sum_{abxy}f_{abxy}M_{a|x}\otimes M_{b|y}-M_{a|\x}\otimes \id$, where $\{M_{a|x}\}$ and $\{M_{b|y}\}$ are valid sets of measurement operators. This in turn is equivalent to  $\mathcal{F}_a\succeq 0$ for all $\mathcal{F}_a=\sum_{abxy}f_{abxy}M_{a|x}\otimes M_{b|y}-M_{a|\x}\otimes \id$.
We say that $\mathcal{F}_a$ admits a sum of squares (SOS) decomposition of degree $2k$, and write $\mathcal{F}_a=SOS_k$ if there exists a set $\{\mathcal{S}^i_a\}$ of polynomials of degree $k$ in the operators $\{M_{a|x}\otimes \id,\,\id\otimes M_{b|y}\}$ such that $\mathcal{F}_a=\sum_i {\mathcal{S}^i_a}^\dagger \mathcal{S}^i_a$ holds for any sets of valid measurement operators $\{M_{a|x}\}$ and $\{M_{b|y}\}$. If this is the case, it clearly follows that $\mathcal{F}_a=\sum_i {\mathcal{S}^i_a}^\dagger \mathcal{S}^i_a\succeq 0$. Therefore, the series of problems
\begin{eqnarray}\label{sdpdual}
G_k(A|E,\x) & =  \displaystyle\min_{\mathbf{f}} &\textbf{f}\cdot\textbf{p}\\
& \mathrm{s.t.} & \mathcal{F}_a=SOS_k\,,\quad a= 1,\ldots,d\,.\nonumber\end{eqnarray}
represents a sequence of relaxations of the dual problem Eq. (\ref{optimdual}) yielding upper-bounds $G_1(A|E,\x)\geq G_2(A|E,\x)\geq \ldots \geq G(A|E,\x)$  on the DIGP.

It is well known that an SOS constraint of the form $\mathcal{F}_a=SOS_k$ can be represented as an SDP constraint \cite{Helton02} and thus that the relaxations Eq. (\ref{sdpdual}) are SDP problems. Such SDP relaxations turn out to be nothing but the dual formulation of the SDP relaxations Eq. (\ref{sdp}) \cite{Navascues08,Doherty08} (see \cite{Pironio10B} for more details on the relation between the primal and dual of the SDP hierarchy).

Even though the primal and dual SDP relaxations Eqs. (\ref{sdp}) and (\ref{sdpdual}) are equivalent, like the original programs, they have different interpretations. Feasible points of the primal programs correspond to decompositions of $\mathbf{p}$ in terms of supra-quantum behaviors in $\mathcal{Q}_k$. They can be understood as characterizing the strategies available to an adversary which is able to prepare supra-quantum behaviors. Such strategies are not necessarily always available in a purely quantum setting and thus the associated values $G_k(A|E,\x)$ represent upper-bounds on the DIGP. The dual programs on the other hand return explicit Bell expressions certifying that the DIGP cannot be higher than a certain value $G_k(A|E,\x)$. Such bounds are valid -- and optimal -- for any strategy in $\mathcal{Q}_k$ and thus are also valid -- though not necessarily optimal -- for any quantum strategy in $\mathcal{Q}$. In other words, the SDP relaxations Eqs. (\ref{sdp}) and (\ref{sdpdual}) not only give a bound on the DIGP, but also return explicit Bell expressions that can be used in any analysis based on a quantitative relation between the amount of Bell violation and randomness, such as in \cite{Masanes11,Pironio13B,Vazirani12B,Pironio10A,Pironio13A,Fehr13,Vazirani12A,Colbeck12,Gallego12}.

\section{Numerical examples}
In this section we  present three numerical examples demonstrating the advantage in taking into account the complete non-local behavior. In the first two examples, we consider a two-input two-output Bell scenario. We introduce the eight parameters $\langle A_x\rangle=\sum_{a=\pm1} a\, p_\mathcal{A}(a|x)$, $\langle B_y\rangle=\sum_{b=\pm1} b\, p_\mathcal{B}(b|y)$, $\langle A_xB_y\rangle=\sum_{a,b=\pm1} ab\, p_\mathcal{AB}(ab|xy)$, where $x,y=1,2$. Their knowledge is equivalent to the knowledge of the complete set of probabilities $p_\mathcal{AB}(ab|xy)$.

\paragraph*{CHSH correlations in the presence of white noise.}
We first consider the randomness that can be extracted from a mixture of maximally violating CHSH correlations plus white noise, i.e. correlations of the form $v\, \mathbf{q}+(1-v)\mathbf{r}$, where $\mathbf{q}$ are the quantum correlations yielding the maximal CHSH violation of $2\sqrt{2}$ and $\mathbf{r}$ denotes completely random correlations for which $p_\mathcal{AB}(ab|xy)=1/4$ for all $a,\,b,\,x$, and $y$. As a function of the ``visibility" $v$ the CHSH violation is thus given by $2\sqrt{2}\,v$. Naively, one would expect that in such a simple example knowledge of the full non-local behavior is of no greater utility than knowledge of the CHSH violation alone. Surprisingly, Figure 1 shows that this is not the case, although the improvement that we get by considering the full non-local behavior is modest.  We have determined numerically the corresponding optimal Bell inequalities as a function of $v$ by solving explicitly the dual programs. We find that these inequalities all have the form 
\begin{equation}\label{chshpn}
f_{11} \langle A_1B_1\rangle+ \langle A_1B_2\rangle+ \langle A_2B_1\rangle-f_{22} \langle A_2B_2\rangle\,,
\end{equation}
where the coefficients $f_{11}$ and $f_{22}$ are given in Figure~2. The case $f_{11}=f_{22}=1$ corresponds to the CHSH inequality and only arises in the case of perfect visibility ($v=1$). This shows that in any real experiment, in which the visibility is necessarily imperfect (i.e. $v<1$), the optimal Bell inequality for randomness certification is not always the CHSH inequality.

\paragraph*{Randomness from partially entangled states.}
In the second example, we consider the  following set of correlations \begin{gather}\label{exmp2}
 \left\langle A_{1}B_{1}\right\rangle =\left\langle A_{1}B_{2}\right\rangle =v\cos\mu,\\ \quad\left\langle A_{2}B_{1}\right\rangle =-\left\langle A_{2}B_{2}\right\rangle =v\sin2\theta\sin\mu\,,\nonumber\\
  \left\langle A_{1}\right\rangle =v\cos2\theta,\; \; \left\langle A_{2}\right\rangle =0,\; \; \left\langle B_{1}\right\rangle =\left\langle B_{2}\right\rangle =v\cos2\theta\cos\mu\,,\nonumber\end{gather}
where $\tan\mu=\sin2\theta$. For $v=1$ these correlations are obtained by measuring a partially entangled state of the form $|\Psi\rangle=\cos\theta\ |00\rangle+\sin\theta\ |11\rangle$ and give rise to a maximal violation of the $I_{1}^{\beta}$ inequality \cite{Acin12}  ($I_1^\beta =I_\mathrm{CHSH}+\beta \langle A_1 \rangle\leq 2 +\beta $) with 
$\beta=2\cos(2\theta)/\sqrt{1+\sin^{2}(2\theta)}$. 
 A value of $v<1$ corresponds to a mixture of these correlations with completely white noise in the respective fractions of $v$ and $1-v$.
 
Figure~3  presents bounds on the global DIGP $G(A,\,B|E,\,2,\,1)$ corresponding to the pair of outcomes associated with the measurements $A_2$ and $B_1$ as a function of $\theta$ for $v=0.99$. We see that taking into account complete sets of correlations can provide a very significant advantage, not only as compared with taking into account only the violation of a single Bell inequality, but also violations of two independent Bell inequalities.

It is interesting to see what the optimal Bell inequalities, obtained via the dual formulation of the SDP programs, look like. The significant advantage obtained in Fig. 2 by taking into account complete data suggests that the corresponding optimal Bell inequalities would be more than mere tweaks of any of the Bell inequalities that have thus far been investigated for the purposes of DIRG (essentially the $I_\alpha^\beta$ inequalities of \cite{Acin12}). This intuition is indeed backed up by the numerics. For example, for $\theta=27\pi/200$ ($G(A,B\mid E,2,1)\simeq0.609$) we obtain  the Bell expression
\begin{gather}
 2.74\left\langle A_{1}B_{1}\right\rangle +2.60\left\langle A_{1}B_{2}\right\rangle +2.35\left\langle A_{2}B_{1}\right\rangle -3.86\left\langle A_{2}B_{2}\right\rangle \\
 +1.36\left\langle A_{1}\right\rangle +1.51\left\langle A_{2}\right\rangle -0.390\left\langle B_{1}\right\rangle +2.05\left\langle B_{2}\right\rangle\,, \nonumber \end{gather}
whose local bound is $8.36$.

\paragraph*{Randomness from entangled qutrits.}
As the last example, we consider the two-input, three-ouput Bell-CGLMP scenario \cite{CGLMP}. Specifically, we consider correlations which violate the CGLMP inequality and which arise by performing the measurements specified in \cite{CGLMP} on the family of states
\begin{equation}\label{qutrit}
\alpha |00\rangle + \sqrt{1-2\alpha^2} |11\rangle + \alpha|22\rangle\,,
\end{equation}
with $0\leq \alpha \leq 1/\sqrt{2}$.
For $\alpha=0$ the state is a product state, for $\alpha=1/\sqrt{3}$ it is a maximally entangled two-qutrit state, while for $\alpha=1/\sqrt{2}$ it is a maximally entangled two-qubit state. For $\alpha\simeq 0.6169$ the CGLMP inequality is maximally violated \cite{durt}, while no violation  is obtained for $\alpha\leq \sqrt{3/22}\simeq 0.3693$ using the set of measurements considered.
Figure~4 presents bounds on the randomness $G(A|E,\,1)$, which can be certified in this scenario, for $\sqrt{3/22}\leq \alpha \leq 1/\sqrt{2}$, taking into account only the CGLMP violation or the full non-local behavior. Unsuprisingly, at the point of maximal violation of the CGLMP inequality, we can certify one trit of randomness, i.e. $G(A|E,\,1)=1/3$. However,  taking into account the complete behavior, a large interval of values of $\alpha$ yields $G(A|E,\,1)=1/3$, including values for which the CGLMP violation is small.  These results have been obtained using the second order relaxation of the SDP hierarchy. The range of values of $\alpha$ for which $G(A|E,\,1)=1/3$ may thus turn out to be larger when going to higher order SDP relaxations or using different measurements from those specified in \cite{CGLMP}. 

\section{Conclusion}
We have shown how the device-independent guessing probability can be evaluated by taking into account in a systematic way the complete non-local behavior characterizing a Bell test and not only the violation of a pre-specified Bell inequality. We have also shown that for any given non-local correlations, there exists an optimal Bell inequality that can certify the maximal amount of randomness compatible with such correlations.
Explicit upper-bounds on the device-independent guessing probability and their associated Bell inequalities can be computed by adapting the SDP hierarchy introduced in \cite{Navascues07}. Low order relaxations, as is often the case with applications of the SDP hierarchy, usually already yield the optimal value of the guessing probability.

Our approach can be straightforwardly adapted to quantify randomness  in purely non-signaling settings (i.e. without requiring the validity of quantum theory). The corresponding programs are simply the analogues of Eqs. (\ref{optim}) and (\ref{optimdual}), where the constraints $\mathbf{\tilde p}^a\in\mathcal{\tilde Q}$ and $p'(a|\x)\leq_\mathcal{Q} \textbf{f}\cdot\textbf{p}'$ are replaced by $\mathbf{\tilde p}^a\in\mathcal{\widetilde{NS}}$ and $p'(a|\x)\leq_\mathcal{NS} \textbf{f}\cdot\textbf{p}'$, respectively, with $\mathcal{NS}$ denoting the set of non-signaling behaviors. Since $\mathcal{NS}$ is entirely characterized by linear constraints (the no-signaling constraints \cite{Barrett05B} and the positivity of probabilities), these programs can be solved using linear programming.

We expect that the tools that we have presented will contribute to advancing our fundamental understanding of the relation between non-locality and randomness, and its cryptographic applications. In particular, the simple examples that we have studied (especially Figures 1, 2, and 4) already yield unexpected results that motivate further investigations. Finally, it would be interesting to understand what is the optimal way to incorporate directly our method in protocols for DIRG and DIQDKD taking into account finite statistics effects.

\paragraph{Note added.} Similar results to our own have been obtained independently and in parallel by J.D. Bancal, L. Sheridan, and V. Scarani \cite{bss}.

\paragraph{Acknowledgments.}
We acknowledge financial support from the European Union under the project QCS, QALGO, DIQIP and from 
the F.R.S.-FNRS under the project DIQIP. S.P. acknowledges support from the Brussels-Capital Region through a BB2B grant. J.S. is charg\'{e} de recherches du F.R.S.-FNRS. O.N.S. acknowledges support from the F.R.S.-FNRS under a grant from the Fonds pour la Formation
\`{a} la Recherche dans l'Industrie et l'Agriculture (F.R.I.A.).
The Matlab toolboxes YALMIP \cite{yalmip} and SeDuMi \cite{sedumi} were used to solve the SDPs giving rise to the figures in Section 6.

\begin{figure}[t]
\center{\includegraphics[scale=0.5]{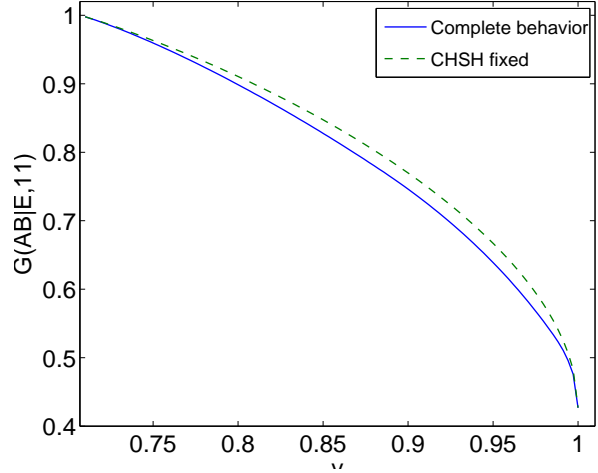}}
\caption{Global randomness $G(A,\,B|E,\,1,\,1)$ as a function of the visibility $v$ for optimally violating CHSH correlations in the presence of white noise. The dashed curve was obtained by taking into account only the CHSH value (i.e. $2\sqrt{2}v$), while the solid curve was obtained by taking into account the full non-local behavior. Both curves were obtained using the second order relaxation of the SDP hierarchy and are actually optimal up to the numerical precision of $10^{-6}$ used (we have verified optimality by finding explicit states and measurements saturating the bounds given by the SDP programs). Except when $v=1$, i.e. when there is no noise, we see that there is a small advantage in taking into account the full non-local behavior.}
\end{figure}
\begin{figure}[t]
\center{\includegraphics[scale=0.5]{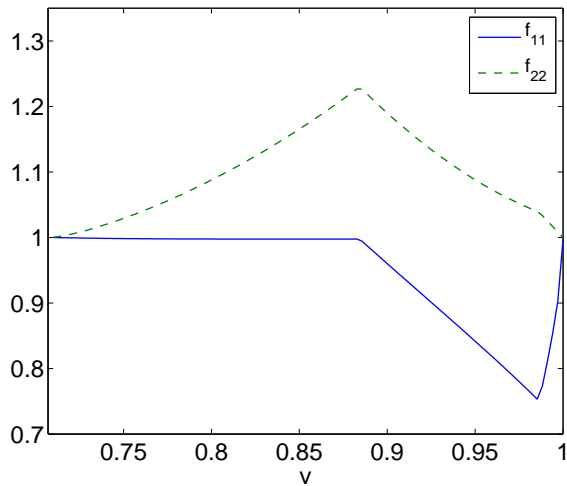}}
\caption{Coefficients of the optimal Bell inequalities Eq. (\ref{chshpn}) as a function of $v$. The CHSH inequality corresponds to the case $f_{11}=f_{22}=1$ and is optimal only for perfect visibility $v=1$ (and trivially $v=1/\sqrt{2}$).}
\end{figure}
\begin{figure}[t]
\center{\includegraphics[scale=0.5]{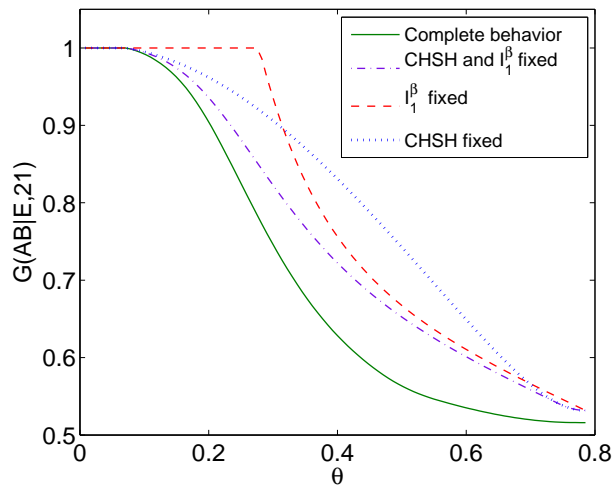}}
\caption{$G(A,\,B|E,\,2,\,1)$ as a function of $\theta$ computed  by taking into account partial or complete non-local data for $v=0.99$. The dashed curve was obtained by constraining only the value of the $I_{1}^{\beta}$ expression, the dotted curve by constraining only the value of the CHSH expression, the dashed-dotted curve by constraining the values of both
$I_{1}^{\beta}$ and the CHSH expressions, and  the solid curve by taking into account the values of all  correlators in accordance with Eq. (\ref{exmp2}). These curves were obtained using the third order  relaxation of the SDP hierarchy. The dashed-dotted curve is optimal up to a precision of $10^{-6}$.}
\end{figure}
\begin{figure}[t]
\center{\includegraphics[scale=0.5]{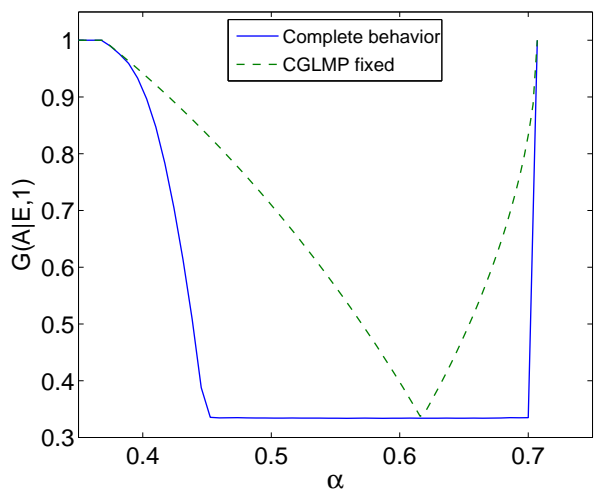}}
\caption{Local DIGP $G(A|E,\,1)$ as a function of the parameter $\alpha$ defined in Eq. (\ref{qutrit}). The dashed curve is obtained by taking into account only the CGLMP value, and the solid one the complete behavior. Both curves were obtained using the second order relaxation of the SDP hierarchy, and the dashed one has been verified to be optimal up to a numerical precision of $10^{-5}$.}
\end{figure}

\end{document}